\documentclass[onecolumn,showpacs,amsmath,amssymb,prd,nofootinbib]{revtex4}
\usepackage{epsfig}

\def\sss{\scriptscriptstyle}
\def\^#1{^{\sss #1}}
\def\_#1{_{\sss #1}}
\def\gmn{g_{\sss\mu \nu}}
\def\Gmn{g^{\sss\mu \nu}}

\def\a{\alpha}
\def\b{\beta}

\def\d{\delta}
\def\m{\mu}
\def\n{\nu}
\def\r{\rho}
\def\l{\lambda}

\def\der#1{{}_{\sss,#1}}

\def\cd#1{{}_{\sss;#1}}
\def\cdu#1{{}_{\sss;}{}^{\sss#1}}

\def\con#1#2{\Gamma^{\sss#1}_{\sss#2}}

\def\f#1{#1-form}

\def\t{\otimes}

\def\ghalf{g^{\sss 1/2}}

\def\div{\vec{\hbox{\mqq\char'162}}\cdot}
\def\grad{\vec{\hbox{\mqq\char'162}}}

\def\div{\vec\nabla\cdot}
\def\grad{\vec\nabla}

\def\sss{\scriptscriptstyle}
\def\^#1{^{\sss #1}}
\def\_#1{_{\sss #1}}
\def\mean#1{\langle #1\rangle}
\def\beq{\begin{equation}}
\def\eeqno#1{\label{#1}\end{equation}}

\def\rarrow{\rightarrow }

\def\dleft{\rlap{{\it D}}\raise 8pt\hbox{$\scriptscriptstyle\Leftarrow$}}
\def\dright{\rlap{{\it
D}}\raise 8pt\hbox{$\scriptscriptstyle\Rightarrow$}}

\def\cmss{{\rm ~cm~s^{-2}}}

\def\az{a\_{0}}

\def\rar{\rightarrow}
\def\s{\sigma}

\def\fm{\varepsilon}
\def\l{\lambda}

\def\Om{\Omega}
\def\f{\phi}
\def\t{\theta}
\def\k{\kappa}

\def\r{\rho}

\def\m{\mu}
\def\n{\nu}

\def\E{\mathcal{E}}
\def\M{\mathcal{M}}
\def\A{\mathcal{A}}

\def\F{\mathcal{F}}

\def\L{\mathcal{L}}

\def\d{\delta}

\def\a{\alpha}
\def\z{\zeta}
\def\xlimin{{x\rarrow\infty \atop{\raise 1pt\hbox to 30pt{\rightarrowfill}}}}
\def\limlim#1#2{{#1\rarrow #2 \atop{\raise 1pt\hbox to 30pt{\rightarrowfill}}}}

\def\vr{{\bf r}}

\def\vv{{\bf v}}

\def\vg{{\bf g}}

\def\va{{\bf a}}
\def\vA{{\bf A}}

\def\vU{{\bf U}}
\def\vX{{\bf X}}
\def\vN{{\bf N}}

\def\grad{\vec\nabla}
\def\div{\vec \nabla\cdot}
\def\gf{\grad\phi}
\def\gxi{\grad\xi}
\def\xiz{\xi\_0}

\def\gN{g\_N}

\def\lM{\ell\_M}
\def\baz{\bar a_0}
\def\haz{\hat a\_{0}}

\def\WW{\mathcal{W}}

\def\A{\mathcal{A}}
\def\azg{\A_0}
\def\dSST{$dS^4~$}

\def\dsr{\ell\_{\Lambda}}
\def\lz{\ell_{0}}
\def\Sz{S_0}
\def\ff{\varphi}

\def\vx{{\bf x}}

\def\hgmn{\hat{g}\_{\m\n}}

\def\alam{a\_\Lambda}
\def\refcite{ \cite}
\def\body{\maketitle}

\begin{document}

\title[MOND from a brane-world picture]{MOND from a brane-world picture\\{\it \small{Dedicated to the memory of Yacov (Jacob) Bekenstein}}\label{ra_ch1}}
\author{Mordehai Milgrom}

\affiliation{Department of Particle Physics and Astrophysics,\\
 Weizmann Institute of Science, Rehovot 76100, Israel }

\begin{abstract}

Yacov (Jacob) Bekenstein contributed greatly to the development of MOND theory. His TeVeS was the bellwether to diverse relativistic-MOND formulations. However, none of the theories we have today is fully satisfactory, and, importantly, we still lack understanding of the roots of the MOND phenomenology. Groping for such understanding I describe a heuristic picture where (nonrelativistic) MOND dynamics emerge in a universe viewed as a nearly spherical brane embedded in a higher-dimensional flat space $\E$.
The brane (centered at $\xi=0$) is described by $\xi(\Om)$; $\xi$ and $\Om$ are the radial and angular coordinates in $\E$.
It couples to a radial potential $\fm(\xi)$ in $\E$ as $\fm\s$; $\s$ is the brane density. The brane tension, $T$, balances the $\fm$ force at radius $\xi=\lz$, implying $\haz\equiv|\fm'(\lz)|\approx T/\s\lz$, yielding for the velocity of small brane perturbations $c^2\approx T/\s\approx \lz\haz$. Local mass density $\r(\Om)\ll\s$ also couples to $\fm$ through $\fm\r$. The radial $\fm'\r$ force induces shallow brane indentations
$\z(\Om)=\xi(\Om)-\lz$, $|\z|\ll\lz$, which, in turn, affect masses in a way that makes $\f\equiv\fm[\xi(\Om)]\approx \f_0+\haz\z$ play the role of gravitational potential. To get MOND dynamics, the brane Lagrangian is {\it assumed} a functional of the local brane orientation, whose normal makes an angle $\t$ with the radial direction, $|\tan\t|=|\grad\z|=|\gf|/\haz$. Thus, some $\az\sim\haz\approx c^2/\lz$ plays the role of the MOND acceleration constant in local gravitational dynamics. Aspects of MOND that may emerge naturally as geometrical properties are:  a. The special role in MOND of an acceleration, $\az$, that marks the transition from dynamics much above $\az$ (large brane slopes) to scale-invariant dynamics much below $\az$ (small slopes). b. The intriguing connection of $\az$ with cosmology. c. The Newtonian limit corresponds to local departure $|\z|\ll\lz$; i.e., $\f-\f_0\sim\az\z\ll\az\lz\sim c^2$ - whereas relativity enters (but not treated here) when $|\z|\not\ll\lz$.
The model also opens new vistas for extension, e.g., it points to possible dependence of $\az$ on $\f$, and to $\az$ losing its status and meaning altogether in the relativistic regime. It also reinforces the view that global (cosmological) and local (e.g. galactic) MOND dynamics have to be understood together as inseparable parts of the same construction. The `old' cosmological-constant catastrophe is obviated because the `cosmological' effects of the large brane energy density, $T$, are cancelled (balanced) by the action of the $\fm$ field. The resulting geometry -- spherical of radius $\lz$ -- when interpreted within general relativity, is characterized by a `cosmological constant' $\Lambda\approx \lz^{-2}$. With a view to generalizing the model to space-time branes I discuss possible connections with the nearly-de-Sitter nature of our Universe.
\end{abstract}


\body

\section{Bekenstein and MOND}
Yacov (Jacob) Bekenstein's contributions to black-hole thermodynamics have established him as a central figure in this field, and have turned out to have even further reaching consequences than initially imagined.
These will surely be covered by other contributions to this volume.
Here I concentrate on Jacob's influential involvement in the MOND program, most notably on his leading role in developing MOND theory.
\par
Not only did Jacob play a major part in the development of MOND, but MOND seems to have played a major part in Jacob's scientific life and career, as we can learn from an interview he gave to ``Scientific American'' in March 2005 on the occasion of his receiving the Israel Prize (http://www.sciam.co.il/archive/archives/2453). Here are some excerpts translated from the Hebrew.

SA: The Israel Prize expresses appreciation of the totality of your scientific work; what do you think is your main achievement?

J: This has two answers. One is how people see me, which is expressed in the Prize citation, concentrating on my work on thermodynamics of black-holes. I started dealing with that in the early (19)70s, and you may say that by the end of the 70s my part in the subject ended.... The second answer is from my own point of view; today I work on other things...  . I think that I am still doing important things, but it takes time until the public, and even important scientists, grasp what you have done.

SA: What are you investigating today?

J: Today, I am trying to change general relativity, still within the principles set by Einstein...in order to
reconcile it with relatively new astronomical phenomena, such as that of the dark mass. ...(DM) is not an explanation but an adjustment to match the facts. In contradistinction, if one accepts... (MOND, which propounds)...that the true theory departs from Newton's theory for small accelerations... (the observations)... follow naturally, as direct consequences of the (MOND) equations.... There is a prescription that works well, MOND, but the reason it works so well is not known. You may say that MOND tells us how the real theory of gravity should look. And this is what I am trying to achieve. I have today a relativistic theory...
(more on MOND and its achievements).... I think that if this theory will work in the end, then this is more important than the thermodynamics of black holes.

Jacob was the first physicist to hear from me about MOND. It was 1982; Jacob was still at Ben Gurion University in Beer Sheba. I went there from Rehovot with my initial MOND trilogy of preprints in my bag to tell Jacob and benefit from his advice.
I had not prepared him for what was in these preprints. Jacob was immediately captured, but he also warned me -- as I vividly remember -- that this is going to encounter much opposition, but also that I should not heed such opposition. He was drawing on his own experience with black-hole entropy and on how his ideas had been received a decade earlier.
Our meeting started a very fruitful and enjoyable collaboration that was to last for 30 years.
\par
The development of MOND paradigm owes much to Jacob. Without doubt, his two peak contributions were the following:
a. His very first contribution to MOND in work we collaborated on, describing the first full-fledged nonrelativistic MOND theory (Jacob dubbed it `AQUAL') \cite{bm84}. This theory was derived from an action with the standard symmetries, and so it enjoyed the standard conservation laws, an important advance on my initial formulation. This work also clarified an important issue concerning the center-of-mass motion of composite bodies.
b. After several attempts to construct a relativistic MOND theory -- initially with me, but then mainly by himself, and notably with Bob Sanders -- Jacob came up with his famous TeVeS theory \cite{bekenstein04}, building on ideas by Bob Sanders (TeVeS sprang from Sanders's `stratified' MOND theory \cite{sanders97}).
\par
The advent of TeVeS has greatly advanced the case for the whole MOND paradigm. Among other things, it attracted to MOND people who were interested in relativity, and as a result several other relativistic MOND formulations were proposed.
\par
About a quarter of Jacob's publications since his first hearing about it were on MOND (about 20 odd out of 80 odd papers).
Jacob also gave numerous talks (e.g., \refcite{bekenstein04a,bekenstein09}), and wrote reviews on MOND (e.g., Refs. \refcite{bekenstein06,bekenstein11a}).
\par
Jacob's general attitude towards MOND had seen ups and downs, steered in part by his own momentary successes and failures in his attempts to advance it.
Following are some examples of excerpts from his letters.

Letter from Sept 8 1985: ``I tried a new approach to Relativistic Mond which is guaranteed to be causal by definition, but the first indication is that it is not giving the right non-relativistic form. Right now I am pretty close to giving up on relativistic theories. The continual failures do not bode well for the whole idea, and I am less than ever interested in going back to empirical tests.''

January 3 1986 (translated from Hebrew); ``Today when I read your letter anew your arguments seem more valid, and I understand less why I thought there are problems. By and large my feeling about MOND is now better....I have a new idea for a relativistic theory....''

August 21 1998 (discussing a colleague he met in France): ``...He knows about
MOND and I explained to him the problems in turning it relativistic.  He
had a suggestion to use a theory  where the physical metric is put together
from the Einstein metric and a $B_{\mu\nu}$ antisymmetric potential.  ... So far as I can see this will also fail to have MOND
behavior....Right now I am exploring an idea where the connection does not come from the
metric but has independent dynamics.  It is slow going.''

Sept 7 2000: ``Yes, one feels the tide will turn; perhaps this is not the beginning of the end of DM but probably the end of the beginning...; just waiting to do some thinking about relat. MOND, which, I think, is becoming urgent.''

Nov 16 2003 while writing the TeVeS paper: ``For a month or so I have been back at writing that paper on
relativistic MOND.  I find, as usually happens, that I did not do all
the needed calculations before I started writing, so progress is slow.
However, I am now at the stage having to do with MOND vs Newtonian
limits, PPN tests and the like, and that is the last of the big issues
in the paper.''

Jacob's influence on me was twofold. I learned much science from him, especially that at the time I knew only little general relativity. I also greatly admired Jacob's knack for presentation and exposition, his clear and precise use of words always impressed me. So much so that in the early days of our collaboration I found myself, many times, trying to emulate his style.

\section{Introduction}
\label{introduction}
Vying with the dark matter paradigm, MOND \cite{milgrom83} contends to account for the mass discrepancies observed in the Universe without dark components. It does so by modifying Newtonian dynamics (ND), and with it, perforce, general relativity (GR). References \refcite{fm12,milgrom14} are recent extensive reviews of MOND.
MOND introduces a new acceleration constant, $\az$, below which dynamics depart from ND and GR. For much higher accelerations; i.e., in the formal limit $\az\rar 0$, a MOND theory should tend to standard dynamics. (In analogy with the `correspondence principle' of quantum mechanics going to the classical for $\hbar\rar 0$.) In the opposite, `deep-MOND limit' (DML) $\az\rar\infty$, keeping $\az G$ fixed, MOND dynamics become space-time scale invariant (SI) \cite{milgrom09}.
\par
There are various `galactic laws' that MOND predicts \cite{milgrom83,fm12,milgrom14a} which have been thoroughly tested and vindicated. For example, a tight relation between the asymptotic rotational speed in a galaxy and its total baryonic mass (tested, e.g., in Ref. \refcite{mcgaugh11}); a general, `mass-discrepancy-acceleration relation' (tested, e.g., in Refs. \refcite{sanders90,mcgaugh04,wk15,mcgaugh16}); and a tight correlation between the central surface density of a (disc) galaxy, and that attributed to a putative dark-matter halo \cite{milgrom09a,milgrom16} (tested, e.g., in Ref. \refcite{lelli16}). MOND predicts detailed rotation curves of individual disc galaxies, given only their baryon distribution (e.g., Ref. \refcite{fm12} and many others).
\par
The MOND constant, $\az$, has been consistently determined from various observed manifestations of these MOND laws (e.g., Refs. \refcite{milgrom88,begeman91,mcgaugh04,li18}). After some initial vacillation in the range $\az\sim (1-2)\times 10^{-8} \cmss$, the value has converged on $\az\approx 1.2\times 10^{-8} \cmss$  (beginning in the early 1990s with Ref. \refcite{begeman91}).
As noted early on (e.g., Refs. \refcite{milgrom83,milgrom89,milgrom99}), $\az$ is near in value to some accelerations of cosmological significance:
\beq \baz \equiv 2\pi \az\approx a\_H(0)\equiv cH_0\approx \alam\equiv c^2/\dsr, \eeqno{coinc}
where $a\_H\equiv cH$ is the acceleration associated with the cosmological expansion rate, $H$ (the Hubble constant), and $a\_H(0)$ is its present value, and  $\dsr=(\Lambda/3)^{-1/2}$ is the radius associated with $\Lambda$ -- the observed equivalent of a cosmological constant.
Thus, the MOND length, $\lM\equiv c^2/\az$, is of the order of the de Sitter radius of our Universe, $\dsr$, or of the present Hubble radius, $\ell\_H\equiv c/H_0$.
This numerical `coincidence', if fundamental, may have far-reaching ramifications (e.g., Ref. \refcite{milgrom15}).
\par
While the salient predictions of MOND follow essentially from only the above basic tenets, we need to dress these tenets with a theory, and -- no less desirable -- to understand the origin of MOND, as all signs point to its being emergent from more fundamental concepts.
\par
Indeed, theoretical endeavors (reviewed in Ref. \refcite{milgrom15}) have proceeded roughly along two avenues.
On one hand, full fledged MOND theories -- both nonrelativistic (NR) and relativistic --   that are derivable from an action principle have been put forth. Such theories are conceptually on a par with ND or GR. They are very useful; for example, for demonstrating various matter-of-principle issues in MOND, or for detailed numerical simulations in MOND, since they capture the essence of MOND, satisfy its basic tenets, share the salient predictions, and obey all the standard conservation laws. Yet, it is quite clear that none of the existing theories can be the final word on MOND.
\par
On the other hand, it has been stressed repeatedly that MOND, as it is now understood, must emerge as an effective, approximate theory from a more basic world picture. Such an underlying theory might tell us why it is an acceleration that marks the boundary between the two regimes of dynamics -- the DML, with its SI, and the high acceleration, ND/GR regime where SI is lost -- and whether the value of $\az$ has a special significance in the wider context, explaining perhaps relations (\ref{coinc}).
\par
Suggestions to derive dynamics as an effective description from a deeper, microscopic stratum go back far in time
even for standard dynamics, such as those attempting to derive Newtonian inertia a-la Mach's principle, or the suggestions that the Einstein-Hilbert action for GR gravity emerges from effects of the quantum vacuum \cite{sakh68}, or from thermodynamical considerations, e.g., Refs. \refcite{jacobson95,padmanabhan10,verlinde11}, to name but a few examples. The continued quest for a satisfactory theory of quantum gravity is also an impetus to dig deeper.
\par
For MOND, the impetus to do so is even stronger. The added stimulus is twofold: 1. The appearance of a new `constant of nature' in MOND -- which may call for a more basic interpretation, and, in particular, its possible relation to the state of the Universe at large pointed to by eq. (\ref{coinc}). 2. The fact that all existing, full-fledged theories interpolate `by hand' from the low- to the high-acceleration regime. Surely this interpolation should result from some more basic theory.
\par
Such a realization has lead to diverse suggestions on how to obtain MOND phenomenology from microscopic considerations. Most of these are listed in Refs. \refcite{milgrom14,milgrom15}. Such ideas, while promising to various degrees, have not materialized into full-fledged theories.
\par
For example, in Ref. \refcite{milgrom99}, I showed how, in the background of a de Sitter (hereafter dS) universe -- such as ours is approximately -- the length $\dsr$ can appear in local dynamics in the guise of an acceleration $\az\sim \alam$ in just the way MOND requires. It was based on the observation that the quantum vacuum might serve as an absolute inertial frame.
Following that, many more suggestions along similar lines have been discussed \cite{pikhitsa10,kt10,ho10,lc11,kk11,klinkhamer12,pa12,pazy13,smolin17,verlinde17}.
\par
Other microscopic approaches aim to reproduce -- at least on galactic scales -- a theory on the lines of the first full-fledged NR MOND theory proposed by Jacob and me in Ref. \refcite{bm84}. There, the gravitational potential, $\f$, is governed by the nonlinear generalization of the Poisson equation
  \beq\div[\mu(|\gf|/\az)\gf]=4\pi G\r(\vx), \eeqno{mondpoiss}
which is gotten from the Lagrangian density $\L\propto \az^2\F[(\gf)^2/\az^2]$ [$\m(x)\equiv\F'(x^2)$] replacing the Poisson Lagrangian $\L\propto (\gf)^2$.
Correspondence with the Newtonian limit dictates $\m(x\rar\infty)\rar 1$, while SI in the DML dictates $\m(x\ll 1)\propto x$.
\par
It is worth noting at this point that the DML of eq. (\ref{mondpoiss}) ($\az\rar\infty,~\azg\equiv\az G$ fixed),
\beq\div(|\gf|\gf)=4\pi \azg\r(\vx), \eeqno{dmlpoiss}
[the equality corresponds to the standard normalization of $\az$ for which $\m(x\ll 1)\approx x$] is not only invariant to space scaling $\vr\rar \l\vr$ -- which follows from the space-time SI of the DML in this time-independent equation -- but has a higher symmetry: it is a conformally invariant in the 3-D Euclidean space \cite{milgrom97}.
With the other, usual, symmetries that this theory enjoys (rotations and translations), its full spatial symmetry group is isomorphic to the isometry group of a dS space-time. This might be another hint to a connection of MOND, as manifested locally, to the state of the Universe at large, perhaps in the framework of a conformal-field-theory-de-Sitter duality.\footnote{QUMOND, another NR modified-gravity MOND version \cite{milgrom10a}, may also serve as a target for microscopic approaches. Its DML, $\Delta\f=\az\^{1/2}\div(|\gN|\^{-1/2}\gN)$, where $\gN$ is the Newtonian acceleration field, is SI but not conformally invariant.}
\par
Attempts to obtain effective (NR) gravitational dynamics described by eq. (\ref{mondpoiss}) usually begin with known physical systems that are governed by a similar equations.
Several such analog systems have been listed in Refs. \refcite{milgrom97,milgrom02,milgrom02a}. For example, nonlinear electrostatics with the dielectric coefficient depending on the electric-field strength (a situation that also occurs in Born-Infeld electrostatics).
Another analog is the (continuity) equation for compressible, irrotational (potential) flows of barotropic fluids. An equation of state $P\propto \r^3$ is needed to get the DML.
\par
So, for example, the microscopic approach to MOND lead by Blanchet \cite{blanchet07,blt09,blh17} builds on the nonlinear dielectric analogy, and involves a gravitationally polarizable, omnipresent medium. Another example is the program lead by Khoury \cite{bk16,khoury16} involving an omnipresent fluid, which reproduces DML phenomenology in its superfluid phase if one assumes that it has a $P\propto \r^3$ equation of state.
\par
The present contribution was inspired by yet another analog problem: The shape of a `membrane', embedded in a (Euclidean) space with codimension one is determined by extremizing  some `energy functional'. In the simple case where the energy of the membrane is proportional to its volume, the shape is determined by an equation of the form (\ref{mondpoiss}), with $\m(x)=(1+x^2)^{-1/2}$, and vanishing right-hand side.
\par
This model does not include the analog of sources, and the form of $\m(x)$ is not what MOND dictates. But, as outlined in Ref.  \refcite{milgrom02a}, it allows for generalizations that do.
Naturally, this approach envisages our Universe as a 3-D spatial membrane (hereafter `brane' as is the common term for higher-than-two-dimensional membrane) embedded in a higher dimensional space -- which I will take to be Euclidean -- where in the NR limit the extra dimension appears in dynamics within the brane as the MOND gravitational potential.
\par
Brane analogs of gravity, and brane representations of our Universe -- so called `brane-world' models -- have been long and amply discussed in the literature. An early brane imagery of gravity is the effective attraction between masses placed on a stretched, elastic membrane placed horizontally in the gravitational field near the Earth surface. The sort of gravity that emerges depends on the `elastic' properties -- the `energy function' of the brane.
\par
In  `brane-world' scenarios, the 3-space (plus 1-time) dimensional arena we shortsightedly perceive as `our Universe' is embedded in a higher dimensional space-time (`the bulk'). While the full dynamics are taking place in the bulk cave, only their `shadow' appears as effective, or emergent, dynamics on the our perceived cave-wall Universe, when we  describe phenomena that do not probe directly the higher dimensions.
\par
Interest in such brane-world pictures has greatly intensified in recent decades, largely due to the advent and developments of string theory, and the realization that it naturally gives rise to such diverse world-models, both because string theory entails higher dimensions and because branes play important roles there.  References \refcite{brax04,ms10} are reviews of such theories, giving some history and extensive references.
Following several predecessors, the brane-world models of Refs. \refcite{rs99,dgp00} have greatly boosted studies of such pictures.
\par
The brane-world models studied to date come in great variety.
They differ, in the first place, in the aspects of the observed Universe they are trying to reproduce. They also differ greatly in the underlying features invoked to such ends. For example, in the nature of the embedding space -- e.g., its dimension, topology, and geometry (i.e., whether it is compactified or not); in the number of branes involved in addition to the one we live on, and their presumed mutual effects; in the (brane and bulk) degrees of freedom involved, and in the interactions and potentials that dictate the dynamics of these; in the initial conditions assumed when time evolution of the model is studied; etc.
\par
By and large, these models, with their various invoked idiosyncracies, do not follow -- certainly not in detail -- from more fundamental theories. Rather, they are largely constructed `by hand' to address various issues or conundrums in particle physics, cosmology, or gravity in general (classical and quantum).
\par
To give but a few examples, the model of Ref. \refcite{rs99} envisages an anti-de-Sitter embedding space and has ramified into one-brane and two-brane classes of models.
Reference \refcite{dgp00} suggested and studied 4-D Gravity on a Brane in 5-D Minkowski Space, and
Ref. \refcite{deffayet01} considered cosmology in this latter picture. Some other models view our Universe as the lowest dimensional echelon in a cascade of several nesting branes \cite{derham08}.
The earlier version of the Ekpyrotic universe \cite{khoury01} aimed to explain aspects of cosmology, otherwise attributed to inflation. It invokes the event that lead to the big bang as a collision of our brane Universe  with another, parallel `bulk brane' that `peeled off' yet another `boundary brane'.
\par
Membrane-induced classical gravity models have also been much discussed extensively by Carter and coworkers, e.g. in Refs. \refcite{carter00,carter01,carter02,uzan02,carter11}.
\par
In my superficial perusal of some of the relevant literature I have not identified elements that might specifically serve in constructing MOND-inspired brane-world models (but see a somewhat related approach in Ref. \refcite{cai18}).
\par
I comment here, as an aside, that BIMOND -- the bimetric formulation of relativistic MOND \cite{milgrom09b}, and other bimetric theories, which involve two metrics defined on a single space-time,
may perhaps be obtained as near-contact (small-distance) approximations of a system of two branes in a bulk, each with its own Einstein-Hilbert action plus a mutual interaction.
\par
Here I greatly expand on the MOND-from-brane idea introduced in Ref. \refcite{milgrom02a}.
A crucial novel feature of the idea is the presence of an external potential $\fm(\vX)$ in the embedding space that acts on brane and matter. The potential is highly symmetric (planar, spherical, etc.).
It defines a local, preferred direction, $\vU$ -- a unit vector in the direction of $\grad\_{\vX}\fm$ -- that plays several roles: 1. $\vU$ points to the direction that appears in the internal brane dynamics as the gravitational potential. 2. In order to obtain MOND dynamics the energy density of the brane has to depend on the local angle between the normal to the brane and $\vU$ (akin to the energy of a dipole layer in an electric field). 3. The field $\vU$ determines the global average geometry of the brane (e.g., a spherically symmetric $\fm$ induces a nearly spherical brane). 4. The MOND constant emerges as the value of $|\grad\_{\vX}\fm|$ at the position of the brane. 5. The same $|\grad\_{\vX}\fm|$ appears in the global balance of the brane at some average radius $\lz$, which leads to $\az\sim c^2/\lz$.
\par
These models, simplistic and heuristic as they are, do suggest possible explanations to -- or at least interpret in geometrical terms -- some important aspects of MOND. For example, the local orientation of the brane with respect to $\vU$ is gravitational acceleration (in units of $\az$). A slope of 1 appears as $\az$; the DML corresponds to near alignment of the local brane normal with $\vU$, while the Newtonian regime corresponds to high inclinations. More can be understood if the brane is a sphere or pseudosphere of
radius $\dsr$ whose shape is locally distorted by masses that cause `indentations' in the brane.
In such a picture NR systems are described by shallow indentations of depth $\ll \dsr$, while highly relativistic gravitational fields correspond to indentations whose depth is $\sim \dsr$. It is also possible to demonstrate why  $\az$, $\dsr$, and the propagation speed of perturbations on the brane (standing for $c$), are related by a relation resembling (\ref{coinc}).
\par
This is only a laying out of the main principles. Practically every aspect of the model I describe requires further elaboration to make the model truly relevant to our Universe.
In Sec. \ref{model}, I start to describe the elements of the model concentrating on local aspects. In Sec. \ref{global}, I consider global aspects, viewing the brane as a distorted sphere.
In Sec. \ref{desitt}, I reconsider the model in the context of a dS universe. Section \ref{discussion} is a discussion.

\section{A brane model for MOND}
\label{model}
One starts with a $d$-dimensional surface, $\M$, embedded in a $d+1$ dimensional Euclidean space, call it $\E$, with Euclidean coordinates
$\vX=(x\^1,...,x\^d,x\^{d+1})$ of the point $\vX$ in $\E$. The brane can be described by specifying one of the coordinates, say $\xi\equiv x\^{d+1}$, as a function, $\xi(\vx)$, of the $d$-dimensional vector $\vx=(x\^1,...,x\^d)$. [The interpretation of $\xi$ and dynamics of the system, described below, ensure that $\xi$ does not become multivalued, as this would lead to infinite gravitational accelerations $\vg\propto\grad_x\xi$; see Sec. \ref{relativity} below.]
The metric induced on the brane from the Euclidean metric of $\E$ is $g\_{ij}=\d\_{ij}+\xi\der{i}\xi\der{j}$ whose determinant is $g=1+(\gxi)^2$  ($\gxi\equiv \grad\_{\vx}\xi$), so the volume of the brane is
\beq V=\int \ghalf d^dx=\int d^dx[1+(\gxi)^2]^{1/2}. \eeqno{vola}
Configurations $\xi(\vx)$ that extremize $V$ satisfy
$\div[\m(|\gxi|)\gxi]=0$, with $\m(x)=(1+x\^2)\^{-1/2}$.
I will want to identify $\xi$ with the NR gravitational potential; so here $V$ may be identified with the action for this potential.
\par
To turn this into an analog for NR MOND gravity as formulated, e.g., by eq. (\ref{mondpoiss}), we need to do at least the following: a. Justify the identification of $\xi$ with the NR gravitational potential, $\f$.  b. Introduce matter. c. Account for the appearance of $\az$, with its different roles, in local dynamics. d. Account for the specific MOND limits: SI in the DML, $|\gf|\ll\az$, and the Newtonian limit for $|\gf|\gg\az$. e. Account for the breakdown of this NR picture beyond the Newtonian and into the relativistic regime. f. Account for the coincidence (\ref{coinc}).
Accounting fully for relativistic phenomena is even a further goal.
I will not be able to achieve even all the goals a-f above, but for some I will indicate possible directions.
\par
In the discussion below I shall refer to the following three standardly used scales: 1. `Global' will refer to the whole brane; e.g., whether it is on average planar or spherical. 2. `Local' refers to local variations in $\xi$ due to the presence of matter.
3. `Average local' which will refer to properties averaged over scales much smaller than the brane's extent, but large enough to homogenize over local roughness of the brane.
Homogeneity and isotropy of the `local average' brane structure are assumed here as an expression of the `cosmological principle'. I shall refer to a local average piece of the brane as $\mean{\M}$.
\par
The action $V$ in eq. (\ref{vola}) is invariant under rotations of the whole brane in $\E$. It thus does not single out a direction the displacement along which can be identified with the potential.
We need to invoke a preferred direction in $\E$ which will determine this direction dynamically via interaction with the brane and with matter. Call the unit vector in this direction at a local position on the brane, $\vU$.
A minimalistic way do this is to assume some potential, $\fm(\vX)$ in $\E$ which couples to the brane and to matter, the local direction of whose gradient is $\vU$.
\par
To get some kind of gravity it is enough for $\fm$ to exert forces on matter (in the $\vU$ direction). This however is not enough to get MOND dynamics.
As explained below, accounting for local MOND gravity {\it requires} that there is also energetic cost to misalignment between $\vU$ and the normal to the brane, $\vN(\vx)$. Namely, the brane energy increases with increasing angle $\psi(\vN,\vU)$. I shall come to this later; here I note that this
has the added benefit that it dynamically makes the brane perpendicular to $\vU$ on local average.
\par
Had this not happened dynamically, we would have had to invoke it by fiat. A component of $\vU$ tangent to $\mean{\M}$ would introduce a preferred direction in the brane that would brake the isotropy of our Universe, breaking the cosmological principle, which we want to retain.
\par
In Sec. \ref{global}, I consider the global structure of $\M$, in particular taking it to be globally a sphere of radius identified with $\dsr$. In this case $\vU$ is radial from some center, and $\fm$ is a function of $|\vX|$.
But, for simplicity of presentation I take for the nonce $\vU$ to be constant; this causes $\M$ to be planar globally and locally on average, which is adequate for considering local dynamics on scales much smaller than $\dsr$.

\subsection{Matter and its Lagrangian\label{inertia}}
`Matter' will be represented my `masses' confined to the brane: either isolated objects characterized by a mass parameter $m$, having the dimensions of inertial mass, or a continuous distribution of such masses of density $\r(\vx)$.
\par
Masses are assumed to have inertia tangent to $\mean{\M}$ -- i.e., perpendicular to $\vU$ --  but not in the $\vU$ direction. Thus, inertia is {\it not} a property in $\E$.
The kinetic Lagrangian of a mass is of the form $(1/2)m\vv^2$, where $\vv=d\vx/dt$ [with no contribution from $(d\xi/dt)^2$].
\par
Had we assumed isotropic inertia in $\E$, the kinetic term for masses would have been $(1/2)m[\vv^2+(d\xi/dt)^2]=(1/2)mv\_i(\d\_{ij}+\xi\der{i}\xi\der{j})v\_j$. As will become clear below, to reproduce MOND gravity we need to consider and account for all values of $\gxi$, large as small. Without the assumption of no inertia in the $\vU$ direction, the wrong inertia will result. I missed this important point in Ref. \refcite{milgrom02a}.
\par
Such confinement of inertia to $\mean{\M}$ may seem natural in a model in which all matter degrees of freedom are confined to $\M$. But note that here inertia is not in the local tangent to $\M$ but in the tangent to $\mean{\M}$, which point to long-rage cause of inertia. I think that all this points to inertia itself being an emergent attribute of matter.
\par
For example, this will come about if inertia is an acquired attribute of masses -- a-la Mach's principle -- due to their interaction with some physical agent that is confined to $\M$ (such as the quantum vacuum \cite{milgrom99}).
\par
Alternatively, absence of inertia in the $\vU$ direction might result if inertia emerges directly from effects associated with the $\fm$ field, which produces a force field in the $\vU$ direction  (somewhat in a similar vein to a magnetic field not impeding motion of charged particles along it, only perpendicular to it). For example, a kinetic Lagrangian of the form  $(m/2)(\dot\vX\times \vU)^2$ gives in the equation of motion the inertial term $m[\vA-(\vA\cdot\vU)\vU]\equiv m\va$, where $\va$ is the projection of the acceleration $\vA$ perpendicular to $\vU$.
\par
Masses are also the couplings of matter to the $\fm$ field, which appears in the Lagrangian as $m\fm(\xi)$.
I take $m$ to be both the coupling strength of masses to $\fm$ and the coefficient of inertia -- an assumption that requires justification. But, a. one needs to assume only proportionality, i.e., that the ratio of the two couplings is the same for all masses, any proportionality constant can be absorbed into $\fm$; b. If inertia too has something to do with the agent that produces $\fm$, as suggested above, such a proportionality is natural.
\par
In summary, the matter degrees of freedom appear in the action as
\beq I\_M=\int d^dx~\r(\vx)\{\frac{1}{2}\vv^2-\fm[\xi(\vx)]\}, \eeqno{matac}
where the density of masses is $\r=dm/d^dx$.\footnote{$\r$ is not defined as the mass per unit volume of the brane
$dm/dV=\r[1+(\grad\_{\vx}\xi)^2]^{-1/2}=\r\cos\psi$, where $\cos\psi=\vN(\vx)\cdot\vU$.}
\par
From the action (\ref{matac}) we get the equation of motion of masses:
\beq \va=\frac{d\vv}{dt}=-\grad_x\fm. \eeqno{eomo}
So clearly, $\f(\vx)\equiv\fm[\xi(\vx)]$ can be interpreted as the gravitational potential for masses on the brane.
\par
According to this picture we sense and interpret the projection of motion tangent to the branes `local average' (here, parallel to the $\vx$ hyperplane) as a change in location. But the motion in the $\xi$ direction is dictated by the $\vx$ motion due to the confinement to the brane, and is perceived as a change in the potential, not as motion. This sensation occurs because our experience tells us that we can move along the $\vx$ hyperspace at will, but we cannot change $\f$ at will, without `moving'; i.e., without changing $\vx$.

\subsection{The brane action}
The brane action translates to the free (kinetic) action of gravity, analogue to the Poisson or the Einstein-Hilbert actions. Here, the discussion is restricted to NR gravity; so masses are assumed to move on the brane at speeds much smaller than the speed of small brane perturbations (see Secs. \ref{global} and \ref{relativity}). We shall naturally identify this latter speed with the speed of gravity (or of gravitational waves). Thus, I consider, at present, only quasi static situations where the brane adjusts its shape instantaneously to the position of the masses: the kinetic action of the brane will thus be neglected. The Lagrangian of a given configuration of the brane is thus (minus) the energy density of the brane configuration, $\xi(\vx)$, which encapsules its `elasticity' and interaction with the $\fm$ field.
\par
I assume a local action so it is $I=\int \L d\^dx$, where $\L$ is some local functional of $\xi(\vx)$.
Without further guidance from a more basic, microscopic, theory of the brane dynamics, the a priori gamut of possibilities for picking $\L$ is very wide. There could be terms that depend on the intrinsic geometry of $\M$, or its extrinsic geometry in $\E$, which would involve second derivatives of $\xi$. This, in itself is not deleterious -- only higher time derivatives lead to Ostrogradsky instabilities -- but I will consider Lagrangian with only first space derivatives of the potential appearing. So in general $\L=\L(\xi,\gxi)$. Requiring rotational symmetry in $\vx$ restricts us to $\L=\L(\xi,|\gxi|)$.
In this preliminary discussion we ignore the interaction of the $\fm$ field with the brane mass analogous to the $\r\fm$ Lagrangian for matter. We shall see in Sec. \ref{global} that this interaction stabilizes the brane at some average position $\xiz$. Given that, we can take here in the planar (local) case $\L$ to be invariant to $\xi$ translations; i.e., $\xi$ independent. We thus end up here with a brane Lagrangian density that is a function of the angle $\psi$ between $\vN$ and the $\xi$ axis (or $\vU$), since $|\gxi|=\tan\psi$.
This is a cornerstone of the present model. It might be likened qualitatively to the effect of an electric field, which applies a moment on a dipole layer forcing it to align perpendicular to the field.
\par
Adding this action to the matter action (\ref{matac}) (taking henceforth $d=3$) we have for the total action
\beq I=-\int F(|\gxi|)d^3x +\int\r(\vx)\{\frac{1}{2}\vv^2-\fm[\xi(\vx)]\}d^3x. \eeqno{memaction}
\par
So far, $\xi$ has been our brane degree of freedom. To reproduce an equation such as (\ref{mondpoiss}) we want $\f$ to be our degree of freedom. A direct translation can be made under the following approximation:
In Sec. \ref{global} it will become clear that the NR limit, which I treat here, corresponds to small departures of $\xi$ from the average $\xiz$ of the brane. `Small' here means much smaller than the extent over which $\fm'(\xi)$ varies appreciably. We can thus take in this NR approximation $\fm(\xi)\approx \fm(\xiz)+\fm'(\xiz)(\xi-\xiz)$.
Defining $\haz\equiv \fm'(\xiz)$ -- a constant with the dimensions of acceleration -- we can write $\gxi\approx\gf/\haz$ in the argument of $F$.
Local shape distortions of $\M$ from the flat are thus perceived as gravitational fields.
\par
The Newtonian limit occurs for steep indentations, $|\gxi|=\tan\psi\gg 1$. The DML corresponds to shallow gradients $|\gxi|=\tan\psi\ll 1$.
\par
For the Newtonian limit, we need $F(Z\rar \infty)\rar K Z^2+const.$, where $K$ sets the general normalization of $F$ and carries its dimensions. In this limit we have
\beq I=-\frac{1}{8\pi G}\int (\gf)^2d^3x +\int\r(\vx)[\frac{1}{2}\vv^2-\f(\vx)]d^3x, \eeqno{newtaction}
where we identified the Newton constant $G$ as
\beq G=\frac{\haz^2}{8\pi K}. \eeqno{ncos}
Define the dimensionless $f(Z)\equiv F(Z)/K$, in terms of which the brane action is $-(\haz^2/8\pi G)\int f(Z)d^3x$.
\par
In the limit of small gradients, which corresponds to the DML, we take guidance from the requirement that the theory be SI. The action for the brane thus has to be invariant to $\vx\rar\l\vx$ when $Z\equiv|\gxi|\ll 1$. A deeper theory, which I do not have, should tell us why this is so; but in default of such a theory we impose this requirement on the brane Lagrangian. The equation of motion of masses being SI tells us that the potential, or $\xi$ is not to change under the scaling operation. The free action for masses is thus scale invariant, and so must then the brane action be in the limit $|\gxi|\ll 1$.
This means that to lowest orders in $Z$ we can write
\beq f(Z\ll 1)= f(0)+ (2/3)\k Z^3, \eeqno{pout}
where $\k$ is a dimensionless parameter, which I assume to be of order unity (see footnote \ref{fafa}). The constant $f(0)$ does not play a role in local dynamics, but it is important in setting the global structure of the brane, since $Kf(0)$ represents the tension, or energy density of the brane unperturbed by masses. Also, note importantly that $\haz^2 f(0)$ plays the role of a cosmological constant, since $Z\ll 1$ almost everywhere in our toy `universe'. We already see that it is of order $\haz^2$, in agreement with the `coincidence' (\ref{coinc}). This will be discussed in Sec. \ref{global}.
The SI simply means that the contribution to the Lagrangian due to the indentations caused by matter is invariant to stretching the brane uniformly (without changing the depth of the indentations).
\par
Defining the MOND constant as\footnote{\label{fafa}This definition of $\az$ corresponds to the standard normalization with which the mass-asymptotic-speed relation reads $MG\az=V^4\_{\infty}$.
It is a convenient normalization when we consider DML physics. $\az$ has also the role of the boundary between the Newtonian and DML regimes at $|\gxi|\sim 1$, or $|\gf|\sim\haz$; here $\haz$ is perhaps a more convenient normalization. An implicit assumption that accompanies MOND's basic tenets is that there are no small dimensionless constants involved, which in particular says that $\az\sim\haz$, or $\k\sim 1$, as assumed.} $\az\equiv\haz/\k$, this limit of the action is
\beq I=-\frac{1}{12\pi \azg}\int [(\gf)^2]^{3/2}d^3x +\int\r(\vx)[\frac{1}{2}\vv^2-\f(\vx)]d^3x, \eeqno{dmlaction}
($\azg\equiv G\az$) and we have in eq. (\ref{dmlaction}) the required DML action leading to the field DML equation (\ref{dmlpoiss}).
\par
More generally, the `interpolating function' appearing in eq. (\ref{mondpoiss}) (whose argument is $|\gf|/\az$) is related to $f(Z)$ by
\beq \m(x)=[f'(Z)/2Z]\_{Z=x/\k}. \eeqno{huta}
$f(Z)$ can be obtained from $\m(x)$ as $f(Z)= 2\int_0^Z Z\mu(\k Z)dZ+f(0)$.
In terms of the equivalent $\n(y)$  interpolating function, also commonly used, [defined by  $x=\k Z=y\n(y)$], we can write $f(Z)=(2/\k) \int ydZ=2\k\^{-2}\{y^2(Z)\n[y(Z)]-\int\^{y(Z)}\n(y')y'dy'\}$, where $y(Z)=\k Z\m(\k Z)$.

\section{Global considerations}
\label{global}
Noting the connection that MOND points to, between local dynamics and global properties of our observable Universe, as in eq. (\ref{coinc}), we need to extend the above planar-brane picture to one with a brane of some characteristic radius. I take it to have a topology of a 3-D sphere embedded in a Euclidean 4-D space, $\E$. The brane geometry is that of an approximate sphere of radius $\lz$ centered at some preferred point in $\E$, distorted by local `indentations' caused by masses. As before, the perturbations on the perfect sphere constitute the gravitational field sensed by us in the brane.
The preferred direction, $\vU$ in $\E$, is now radial from some center, and $\xi$ is the radial coordinate measured from this center.
\par
If $(\xi,\chi,\t,\ff)$ are spherical coordinates in $\E$ with respect to this center, the brane is described by
\beq \xi(\Om)=\xi(\chi,\t,\ff)=\lz+\z(\Om), \eeqno{juya}
with $\z$ having the same sign everywhere.\footnote{Here I shall confine myself to perturbations $|\z|\ll\lz$ -- which we shall see correspond to the NR limit -- so the sign of $\z$ need not be specified. It will become important when we will want to consider, in the future, $|\z|\not\ll\lz$.}
\par
Already at the outset, this picture raises several important questions:
\begin{enumerate}
\item
Since we are seeking to account for a relation of the form $\haz\sim c^2/\lz$, what is the interpretation of $c$, and how is it brought into the picture?

I shall argue that in the model, $c$ is naturally to be identified with the speed of small perturbations on the brane. It enters as the (square root of the) brane's tension-to-density ratio, which is determined -- for a balanced brane -- by the supporting force-per-unit-mass ($\sim\haz$), and by the global radius, $\lz$.
\item
In the planar case, the only `constant of nature' that appears besides $G$ is $\haz$ and the only transition we had to consider is from small to large gradients, $\gxi=\gf/\haz$. Now there is an additional `constant', $\lz$. In most of the discussion I will assume $|\z(\Om)|\ll\lz$, namely local gravity constitutes small perturbation on the global spherical geometry of the brane. But what is the translation of this condition into in-brane dynamics, and what is the meaning of $\z\not\ll\lz$, where one would clearly need to modify the treatment?

I will argue that $|\z(\Om)|\ll\lz$ corresponds to the NR regime $\f\ll c^2$, while $\z\not\ll\lz$ takes us into the regime of relativistic gravity. This follows naturally once we have made the connection $\haz\sim c^2/\lz$, since $\f=\haz \z$.
\item
In the planar case we could ignore the global brane dynamics. But now we have to account for the external forces on the brane itself. If the brane is assumed static (constant $\lz$) on average, what balances it? And if $\lz$ is time dependent, what is the dynamics of these changes.
\item\label{variablea0}
In the planar case we do not have a natural length scale for variation of $\fm(\xi)$, $\fm'(\xi)$, etc. Here $\lz$ provides such a scale. This could open whole new avenues for extending MOND as we now know it. For example, $\xi$ dependence of $\fm'$ translates effectively (but not quite) into dependence of $\az$ on the potential.
\par
Jacob proposed at some point \cite{bekenstein11} that letting $\az$ depend on the local gravitational potential might account for the galaxy-cluster conundrum of MOND; and this proposal was studied in more detail in Ref. \refcite{zf12}. But there $\az$ is required to change substantially within a potential range $\ll c^2$. Here, if $\lz$ is the only scale length that appears, I speak of effective changes in $\az$ from the NR to the relativistic regime, or, more accurately, of $\az$ losing its role and meaning altogether in the relativistic regime. But then again, if $\fm'(\xi)$ does change appreciably with $\xi$ near $\xi=\lz$ -- which introduces another length scale $\ell_1\ll\lz$ into the model -- then $\az$ could depend strongly on the potential over potential changes of order $(\ell_1/\lz)c^2$.\footnote{Expand $\fm'(\xi)\approx \fm'(\lz)+\fm''(\lz)\z\approx \fm'(\lz)+[\fm''(\lz)/\fm'(\lz)]\f$, and define $\ell_1=[\fm''(\lz)/\fm'(\lz)]^{-1}$. Then $\fm'(\xi)\approx \lz^{-1}c^2(1+\lz\f/\ell_1 c^2)$.} Without the geometrical underlying picture, such a dependence would seem rather ad-hoc and unnatural. Here this is quite natural since even adding a constant to $\f$ in some region of space, is tantamount to pushing that region of the brane to another value of $\xi$, where $\az\propto\fm'(\xi)$ may be different.
\item
\label{noso}
Equation (\ref{mondpoiss}) has no solution on a boundaryless brane -- such as treated here -- unless the total source vanishes. So, if eq. (\ref{mondpoiss}) is to emerge, how can this condition be satisfied?
\par
We shall see that the elements required for brane balance introduce terms to the source of eq. (\ref{mondpoiss}) that automatically annul the total source.
This shows that the global dynamics of the brane must be accounted for even when we speak of local dynamics.
It thus supports the claimed made repeatedly in the context of MOND that cosmology and local MOND dynamics must be understood as two aspects of the same construct (as opposed to first constructing a local MOND theory from which cosmology will also emerge).
\end{enumerate}
\par
The arena where we describe dynamics is the sphere $\Sz$ of radius $\lz$, where a convenient choice of coordinates
is $\Om=\chi,\t,\ff$. But we will want to write the dynamics in covariant form (with respect to the $\Sz$ geometry), so that they will automatically reduce to the planar case for local dynamics.
\par
The Euclidean geometry of $\E$ induces the metric $\gmn$ on $\Sz$, and  the metric $\hgmn$ on $\M$. In the
 $(\chi,\t,\ff)$ coordinates, the metric determinants are respectively
\beq \ghalf =\lz^3\sin\^{2}\chi\sin\t; ~~~~~ \hat g\^{1/2} =\xi^3\sin\^{2}\chi\sin\t[1+\eta\der{\chi}\^2+\eta\der{\t}\^2/\sin\^2\chi+
\eta\der{\ff}\^2/\sin\^2\chi\sin\^2\t]^{1/2},  \eeqno{sigthr}
where $\eta\equiv \ln{\xi}$.
The volume element on $\Sz$ is $dV=\ghalf d\chi d\t d\ff=\lz^3d\Om$.
We can write the ratio of the volume element on $\M$ to its projection on $\Sz$ as
\beq (\hat g/g)\^{1/2}=(\xi/\lz)^3[1+(\lz/\xi)^2\xi\der{\m}\Gmn\xi\der{\n}]^{1/2}.  \eeqno{sigthras}
If we put $\lz=\xi$, the local value of $\xi$ at some point, this ratio is the local value of  $1/\cos{\psi}$, where $\psi$ is the angle between the normal to the brane and the radial direction. So,
\beq \cos{\psi}=[1+\eta\der{\chi}\^2+\eta\der{\t}\^2/\sin\^2\chi+
\eta\der{\ff}\^2/\sin\^2\chi\sin\^2\t]^{-1/2}, \eeqno{becost}
Or,
\beq \cos{\psi}=(1+Z^2)^{-1/2}; ~~~~~\tan{\psi}=Z; ~~~~~Z\equiv (\lz/\xi)(\z\der{\m}\Gmn\z\der{\n})^{1/2}. \eeqno{becostI}
Under the assumption $|\z(\Om)|\ll\lz$, which we restrict ourselves to, we can write to lowest order in $\z/\lz$
\beq Z=(\z\der{\m}\Gmn\z\der{\n})^{1/2}. \eeqno{bopuit}
Similar to the planar case, we want an equation from which we can determine the `gravitational potential', $\xi(\Om)$,  and that locally will be of the form of eq. (\ref{mondpoiss}).  We want the effective action
for $\xi$ and its field equation to be covariant with regard to the geometry of $\Sz$. In other words, we want the field equation to be of the form
\beq [\m(Z)\z\cdu{\m}]\cd{\m}=4\pi GS, \eeqno{covar}
where all the geometrical operations (covariant derivatives, raising of indices, etc.) are for $\gmn$.
\par
An important issue immediately arises as alluded to in point \ref{noso} above:
An equation such as eq. (\ref{covar}) does not have solutions in a space with the topology of a sphere,  unless its sources total to zero:
Applying Gauss's theorem to eq. (\ref{covar}), the volume integral of the left hand side gives a boundary term, which vanishes because $\Sz$ has no boundary. Thus it follows that the total source $\int\_{\Sz}S~dV=0$.
\par
The source term of this equation, $S$, would have to be, for some reason, not proportional to the local mass density $\r$, but to some $\r+s$ with $\int\_{\Sz}(\r+s) dV=0$. But, since we want to retain $\r$ as the local source of gravity to a good approximation we will have $s$ itself nearly homogeneous. This will automatically introduce global quantities into the source of our field equation: a crucial element of the model. This will come about when we include in the model the global balance of the brane.
\subsection{\label{soap}Bubble universe}
To the geometry as described above we add the following physics. The brane and matter on the brane are subject to a central potential field $\fm(\xi)$, each couples to it with its own density.

\subsubsection{The matter action}
For the reasons listed above, we assume, as before, that only motion tangent to $\Sz$ is subject to inertia, and enter the kinetic Lagrangian of matter; motion in the $\xi$ direction is not associated with inertia.

The matter degrees of freedom, $\vx$ appear in the action as
\beq \int dV \r\left\{\frac{1}{2}\left[\frac{\xi(\vx)}{\lz}\right]^2\vv\^2-\fm[\xi(\vx)]\right\}, \eeqno{luopinI}
$\r$ is the mass per unit volume on $\Sz$. The velocity that enters inertia is taken as $(\xi/\lz)\vv$, where $\vv$ is the velocity of the projection of the body's position (at radius $\xi$) on $\Sz$.
\par
The matter equation of motion is thus
\beq \left(\frac{\xi}{\lz}\right)^2\dot\vv=-\fm'(\xi)\gxi-\frac{2\xi}{\lz^2}\vv(\vv\cdot\gxi)+\frac{\xi}{\lz^2}\vv\^2\gxi. \eeqno{juert}
Restricting to systems with $\z\ll\lz$ -- which is tantamount to the NR limit -- we take the lowest order in $\z/\lz$. So, as before, we define $\haz\equiv\fm'(\lz)$, $\f\equiv \haz\z$, and put $\xi/\lz\approx 1$ to get
\beq \dot\vv=-\gf-\frac{2}{\haz\lz}\vv(\vv\cdot\gf)+\frac{1}{\haz\lz}\vv\^2\gf. \eeqno{juertus}
$\haz$ plays the role of the MOND constant, $\haz\sim\az$. In the real world, and as we shall see in the model too, $\haz\sim c^2/\lz$, where $c$ is the propagation speed of small perturbations on the brane, with the role of the speed of light. Thus, the last two terms on the right-hand side of eq. (\ref{juertus}) are suppressed by a factor $\sim v^2/c^2$ compared with the first and can be neglected in our approximation ($v^2/c^2\sim\f/c^2\sim\z/\lz\ll 1$) to give
\beq \dot\vv=-\gf. \eeqno{juertusa}
\subsubsection{The brane action}
The action terms that involve the brane's degree of freedom $\xi(\Om)$ and determine its shape are
 \beq I=\int\{ F(\xi,Z)+\fm(\xi)[\r(\Om)+\s]\}dV.  \eeqno{bubbleaction}
Here, $F$ depends on the shape of the brane and {\it on its local orientation with respect to the external, preferred direction} $\vU$. The angle, $\psi$, between the local brane normal and $\vU$ being $\tan{\psi}=Z=(\lz/\xi)(\xi\der{\m}\Gmn\xi\der{\n})^{1/2}$. For example, if this Lagrangian density is proportional the volume element of the brane $F(\xi,Z)=q(\xi/\lz)^3(1+Z^2)^{1/2}$.  The third term is the interaction of the brane with the external potential; $\s$ is assumed constant and can be interpreted as the mass density of the brane (again, per unit volume on $S_0$).
\par
Extremizing this action over $\xi$ gives:
\beq ( \lz^2\xi\^{-2}Z^{-1}F\der{Z}\xi\cdu{\m})\cd{\m}=g\^{-1/2}(\ghalf  \lz^2\xi\^{-2}Z^{-1}F\der{Z}\Gmn\xi\der{\m})\der{\n}=
 \fm'(\r+\s)+F\der{\xi}-Z\xi^{-1}F\der{Z}. \eeqno{bubalcov}
\par
Applying Gauss's theorem now requires not that the total mass vanish, but that
\beq \int\_{\Sz}dV [\fm'(\r+\s)+F\der{\xi}-Z\xi^{-1}F\der{Z}]=0. \eeqno{junfa}
There is, in fact, a continuum of integral relations that solutions obey, namely the general higher-moments `virial relations' for action-governed theories \cite{milgrom94}: Multiply the field equation (\ref{bubalcov}) by $\ghalf \xi^\a$ and integrate over $\Sz$ to give
 \beq \int\_{\Sz}dV \xi\^\a[\fm'(\r+\s)+F\der{\xi}+(\a-1)Z\xi^{-1}F\der{Z}]=0.  \eeqno{viriav}
  The zero-source condition (\ref{junfa}) is the $\a=0$ case.
\par
Condition (\ref{junfa}) can be used to eliminate the `unobservable' $\s$ and write the equation in terms of only the quantities that are measurable
by us in $\M$, viz. $\xi$ (or $\f$) and $\r$. This is done at the cost of ending up with a nonlocal theory, since the right-hand side of eq. (\ref{bubalcov}) becomes
\beq S=\fm'\r-\mean{\fm'\r}^*+F\der{\xi}-Z\xi^{-1}F\der{Z}-\mean{F\der{\xi}-Z\xi^{-1}F\der{Z}}^*, \eeqno{masdas}
where for a quantity $Q$
\beq \mean{Q}^*\equiv \frac{\int QdV }{{\fm'}\^{-1}\int \fm'dV }.  \eeqno{kioyt}
While $\mean{Q}^*$ is not quite $\mean{Q}$, the volume-weighted average of $Q$, as it depends on $\xi$ through $\fm'(\xi)$, we do have $\int\mean{Q}^*dV=\int Q~dV$.
\par
For a constant $\r(\Om)=\r\_0$ the solution is $\xi=\xiz=const.$, and taking $\lz=\xiz$,  eq. (\ref{junfa}) gives the equilibrium condition for the spherical bubble which determines $\lz$:
\beq \fm'(\lz)(\r\_0+\s)+F\der{\xi}(\lz,0)= 0. \eeqno{kalai}
\par
Restricting again to the `nonrelativistic' case (see Sec. \ref{relativity}),with $\fm'\approx\fm'(\lz)\equiv\haz$; also $\mean{Q}^*\approx \mean{Q}$ is a constant, and $Z\approx(\z\der{\m}\Gmn\z\der{\n})^{1/2}$. Then write eq. (\ref{bubalcov}) as
\beq [Z^{-1}F\der{Z}(\lz,Z)\z\cdu{\m}]\cd{\m}=
 \haz(\r-\mean{\r})+\lz\^{-1}[F(\hat F_{\xi}-\hat F_{Z})-\mean{ F(\hat F_{\xi}-\hat F_{Z})}]\_{\xi=\lz}, \eeqno{bubafa}
where  $\hat F_{\xi}\equiv \partial\ln{F}/\partial\ln{\xi}$, and
$\hat F_{Z}\equiv \partial\ln{F}/\partial\ln{Z}$.
\par
To get the Newtonian limit for $Z\gg 1$ we have to have in this limit
\beq F(\xi,Z)\rar K(\xi)Z^2+\bar F(\xi). \eeqno{newlim}
$\bar F(\xi)$ does not appear in the NR field equation.
 Define then, generalizing the planar case, the dimensionless
\beq f(\xi,Z)\equiv F(\xi,Z)/K(\xi),   \eeqno{lioyt}
and $G\equiv \haz\^2/8\pi K(\lz)$,  $\f=\haz\z$, and write eq.(\ref{bubafa}) as
\beq (4\pi G)^{-1}[\m(Z)\f\cdu{\m}]\cd{\m}=
 \r-\mean{\r}+\haz\^{-1}\lz\^{-1}[F(\hat F_{\xi}-\hat F_{Z})-\mean{ F(\hat F_{\xi}-\hat F_{Z})}]\_{\xi=\lz}, \eeqno{bubiop}
 where $\m(Z)\equiv (2Z)^{-1}f\der{Z}(\lz,Z)$, and $Z=(\f\der{\m}\Gmn\f\der{\n})^{1/2}/\haz$.
\par
To get the required SI of the deep-MOND limit we want
\beq f(\xi,Z\ll 1)\approx f_0(\xi)+(2/3)\k(\xi)Z^3, \eeqno{libut}
and define the phenomenological MOND acceleration as $\az=\haz/\k(\lz)$, so that
in this limit $\m=(\f\der{\m}\Gmn\f\der{\n})^{1/2}/\az$.
\par
It will be a great advance if we will be able to understand from the microscopics of the brane why its energy function enjoys such SI in this limit, because this would essentially determine the deep-MOND limit from first principles.
\par
It can be shown that in the nonrelativistic case I consider here, the terms other than $\r$ in source on the right-hand side of eq. (\ref{bubiop}) are suppressed relative to $\r$ near local masses (such as a galaxy, a star, etc.), as it should if we are to reproduce a MOND equation,  $[\m(Z)\f\cdu{\m}]\cd{\m}= 4\pi G\r$. This can be seen as follows:
First, near matter that is highly clumped, clearly $\mean{\r}\ll\r$, and we can neglect $\mean{\r}$ in local dynamics. Second, the contribution to the source of the terms other than $\r$ is suppressed by a factor of order $\z/\lz$ relative to $\r$. To see this, write $F(\hat F_{\xi}-\hat F_{Z})$ as its mean $\mean{F(\hat F_{\xi}-\hat F_{Z})}$, which can be very large (possibly $\gg\r$), plus a position dependent contribution. The first cancels in the source term. The position-dependent part comes from regions where $Z$ and $\xi$ vary due to the presence of nearby masses.  $\hat F_{\xi}$ and $\hat F_{Z}$ are of order unity or less [see, e.g. eqs. (\ref{newlim})(\ref{libut})]; so consider only the variation in $F$, which has contributions $\d F\sim F\der{Z}Z$ and $\d F\sim F\der{\xi}\z$.
Take an isolated, nonrelativistic mass of density distribution $\r$, which gives rise to a potential $\f=\haz\z$.
Multiply eq. (\ref{junfa}) by $\lz$ and subtract from eq. (\ref{viriav}) for $\a=1$; then eliminate $\s$ to get
 \beq \int dV ZF\der{Z}=-\int dV\z\{\fm' \r-\mean{\fm\r}^*+\xi\^{-1}[F(\hat F_{\xi}-\hat F_{Z})-\mean{ F(\hat F_{\xi}-\hat F_{Z})}^*]\}.   \eeqno{jatruf}
Thus the contribution of the extra source terms to the `mass' in eq. (\ref{bubiop}), which is $\sim(\lz\haz)\^{-1}\int dV ZF\der{Z}$ is $\sim (\z/\lz)\int dV\r$, and can be neglected in our approximation.
The same can similarly be shown for $F\der{\xi}\z$.
\par
In summary, if the brane energy density has the appropriate dependence on the brane's shape and orientation -- for microscopic reasons that I cannot pinpoint at present -- NR MOND gravitational dynamics can emerge.
The resulting equation is, strictly speaking, nonlocal, in that the source term involves global integrals over the whole brane. This is required by the structure of the field equation, but is also justified by the dynamics of the model requiring global near equilibrium between the brane `elasticity' with the external force on the brane due to the $\fm$ field.
\par
The appearance of global contributions to the source is such that they nearly annul any position-independent contribution to the source. This may offer a possibility for solving the `old' cosmological-constant (CC) problem; viz., understanding why the vacuum energy acting as a CC does not seem to contribute to the energy content that enters cosmological evolution. It has been suggested before that by modifying GR to become nonlocal (e.g. by introducing into the action integrals over the whole of space-time) this problem can be solved (see, e.g., Refs. \refcite{linde88,tseytlin91,dgs02,ah02,davidson09,gabadadze14}). Interestingly, some of these, such as Refs. \refcite{dgs02,gabadadze14} propose to usher in such nonlocal modifications starting from local actions in some brane-world picture, as here.
Here, such nonlocal modifications are, furthermore, tied with MOND in a way that reproduces naturally the MOND cosmological coincidence, eq. (\ref{coinc}).

\subsubsection{\label{relativity}Meaning of the relativistic limit}
Along our treatment of the brane dynamics we have assumed in many places that we are dealing with the NR limit.
This entered, e.g., when we assumed that masses move much slower than brane perturbations; when we took everywhere
$\z\ll\lz$, or $\f\ll\lz\haz$, and claimed that this is tantamount to $\f\ll c^2$; where we neglected terms suppressed by $v^2/\lz\haz$ [e.g., going from eq. (\ref{juertus}) to eq. (\ref{juertusa})], claiming that this is the same as $v^2/c^2\ll 1$; where we neglected all the source terms in eq. (\ref{bubiop}) except $\r$ because they are suppressed by $\z/\lz$, etc. Our treatment clearly breaks down for $\z\not \ll \lz$.
But why are all the above approximations valid in the NR limit? Clearly, if it is correct that $\lz\haz\sim c^2$, where $c$ is the speed of small brane perturbations, this interpretation is valid. With the values that $\haz\sim\az$ and $\lz\sim\dsr$ attain in the real Universe this is clearly so due to relation (\ref{coinc}).
But does the model, self consistently imply that $\lz\haz\sim c^2$?
\par
So far we have not discussed the kinetics of the brane: we have had no kinetic term in the brane action, assuming that the brane adjusts its shape instantaneously to the configuration of matter.
But, if $\s$, the brane coupling to the potential $\fm$ can also be identified with its inertial-mass density, as was the case for matter, then the equilibrium condition (\ref{junfa}) says that $\haz\lz$ is the ratio of the brane tension to its mass density.\footnote{I am neglecting here the contribution of matter to the total force on the brane, making the plausible assumption that $\s\gg\mean{\r}$. (This is certainly true if $\s$ has a large contribution from vacuum energy density.)}
This is indeed the square of the velocity of propagation of small perturbations of the brane's shape, which we can identify with $c^2$ of our toy model.
\par
{\it The intriguing coincidence (\ref{coinc}) thus finds a natural origin in our model.}
\par
Finite brane inertia leading to a finite speed of propagation of perturbations also affects the matter free action.
Space does not allow to treat this here, but it may well be that (relativistic) inertia of matter can emerge from the matter-brane interaction that we already have in the model, together with brane inertia: Energetic cost on motion of masses may come from the motion of the brane induced by matter motion. This is analogue to the mechanisms described in Ref. \refcite{milgrom06} where bodies move in a compressible fluid with which they interact. There are, in fact many instances of emergent-inertia scenarios of bodies moving in a medium with which they interact.\footnote{For example, the effective mass of electrons in solids, Euler-Heisenberg correction to the Maxwell action due to vacuum effects, etc.}
\par
Note in this connection that the role of $c$ as a limiting speed of masses on the brane emerges automatically: Motion with $v>c$ leads to `shock' formation of the brane with infinite values of $\gf$, and so infinite brane energy densities.
\par
There are no systems that are both relativistic $\z\sim \lz$ and in the DML $\gxi\ll 1$ because this would mean that the extent of the system is $\gg\lz$ which is not possible. This is the model geometrical meaning of the observation that in the real Universe we cannot have systems with $\f\sim c^2$ and $|\gf|\ll\az$. For example we cannot have DML black holes because this would require their Schwarzschild radius to be $\gg\dsr$.
\par
Because relativistic gravity is described by many degrees of freedom (e.g., the metric components and possibly additional fields) extending the model to the full relativistic regime may require embedding in a space of higher codimension, with enough extra dimensions to account for the many gravitational degrees of freedom. For example, accommodating six gauge-fixed components of a 4-D metric on a 4-D brane would require a ten-dimensional embedding space.
\par
We saw that $\haz$ enters NR local dynamics as $\fm'(\lz)$, applicable for $\z\ll\lz$. But clearly
$\haz$ (or $\az$) will lose its status in the extreme-relativistic domain. As alluded to in point (\ref{variablea0}) above, to  first order we can see that $\haz$ has to be replaced by $\fm'(\xi)\approx\haz+\fm''(\lz)\z=\haz[1+p\f/c^2]$, where $p\equiv (d\ln{\fm'}/d\ln{\xi})_{\lz}$ is presumably of order 1 (unless $\fm'$ varies quickly around $\lz$). This introduces `relativistic' corrections to our preceding results (together with other corrections of the same order). This cannot be described simply as MOND with a potential-dependent $\az$ (for example because $\haz$ also entered the definition of $G$).
\par
But clearly, in the relativistic regime $\az$ loses its significance altogether -- a possibility discussed oftentimes (see e.g., Ref. \refcite{milgrom15}). Various existing relativistic extension of MOND can be described as a covariant theory of gravity that hinges on an acceleration constant $\az$, tends to GR in the limit $\az\rar 0$, and satisfies the MOND basic tenets in the NR regime. Our model points to the possibility that this is not the correct type of theory.
\par
\begin{figure}[h]
	\centering
\includegraphics[width=0.7\columnwidth]{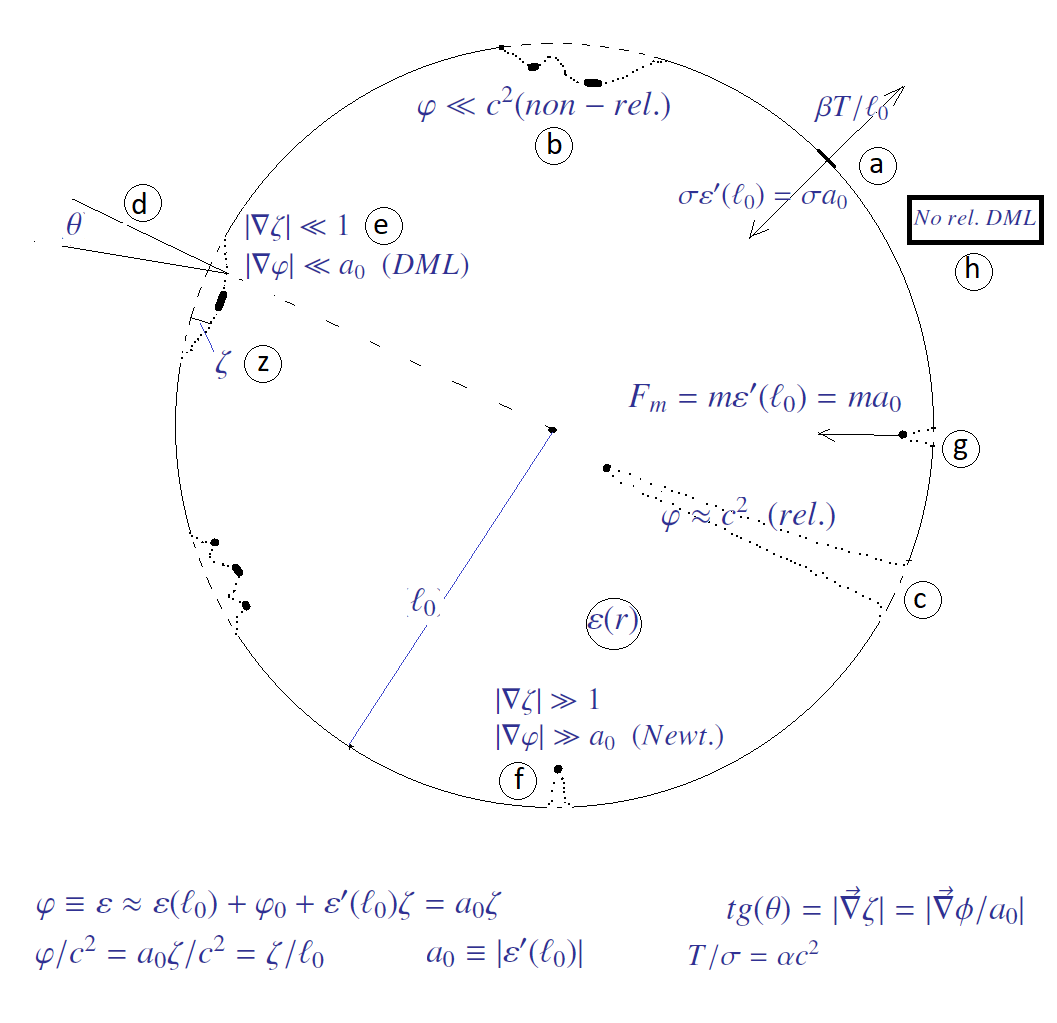}
\caption{Schematics of the brane dynamics: The brane (of density $\s$) and `masses' ($m$)  on it couple to a spherical potential $\fm$ in the embedding space (encircled $\fm$ in the Figure). The brane thus attains an on-average spherical shape of radius $\lz$, for which the force due to brane pressure (tension), $T$, which per unit volume is  $\b T/\lz$ ($\b$ is a geometrical factor of order unity) is balanced by the force per unit volume $\s\fm'(\lz)\equiv\s\az$ (encircled ``a'' in the Figure).
The speed of propagation of perturbations on the brane is $c^2=\a^{-1}T/\s=(\b\a)^{-1}\lz\az$ ($\a$ of order unity).
The nonrelaivistic gravitational potential $\f\equiv \az\z$, where $\z$ is the local departure from sphericity (encircled ``z'').
A shallow depression $\z\ll\lz$, which is equivalent to $\f\ll c^2$, characterizes nonrelativistic gravity (encircled ``b''), an approximation that is broken  where $\z\not\ll\lz$ (encircled ``c''). $\t$ is the angle between the normal to the brane and the radial direction (encircled ``d''). The limit $|\tan\t|\ll 1$ -- the same as $|\grad\z|\ll 1$, and the same as $|\gf|\ll\az$ -- corresponds to the deep-MOND limit (encircled ``e''), while $|\tan\t|\gg 1$ is the Newtonian limit (encircled ``f'').
The $\fm$ force on a mass $m$ is $m\fm'$, which for `nonrelativistic' depressions is $\approx m\fm'(\lz)\approx m\az$ (encircled ``g''). This breaks down when $\f\not\ll c^2$. A depression that is both relativistic ($\z\not \ll 1$), and in the deep-MOND limit ($|\grad\z|\ll 1$) is not possible, as it would have to have an extent $\gg \lz$ (encircled ``h'').}		\label{figure}
\end{figure}

The general schematics and various relations that underlie the brane model are shown schematically in Figure \ref{figure}.

\section{The de Sitter context\label{desitt}}
Nonrelativistic dynamics addressed in the preceding sections can clearly not remain the last word. It is desirable to extend the picture to a model of our Universe as a $4-D$ space-time (ST) brane embedded, for example, in a higher-dimension Minkowski space-time. As mentioned in the introduction there are various such models in the literature proposed to various ends, but not, to my knowledge, ones that are directly relevant to MOND.
\par
As also mentioned in the introduction, MOND seems to have some affinities with the nearly $4-D$ dS (hereafter \dSST) geometry of the Universe.
Unlike a \dSST, our Universe was characterized by a `big bang', and its dynamics in its cosmological past was dominated by matter. So our complete ST is certainly not \dSST. However, today it appears to be described nearly as a \dSST (its energy density being dominated by a CC-like entity) and it will become even more so in the future. So as a first step it would be interesting to generalize the discussion in the preceding sections to a complete \dSST, as this might give relevant insights regarding our less symmetric Universe.
I leave such idealized extensions to future work. Here I only point some relevant properties of \dSST beyond those already discussed in the introduction.
\par
A \dSST is a maximally symmetric ST with Minkowskian signature and positive curvature. It can be described as a pseudo $4-D$ sphere, embedded in a 5-D Minkowski  ($M^5$) ST, with
coordinates $X\^A$, $A=0,1,...,4$. The collection of all points satisfying (in this section I take $c=1$)
\beq X^2\equiv X\^A\eta\_{AB}X\^B=\dsr^2, \eeqno{iias}
with the induced metric is a \dSST space-time, of radius $\dsr$ [$\eta\_{AB}=diag(-1,1,1,1,1)$.
\par
The $M^5$ coordinates $X\^A(\tau)$ of world line, $\WW$, in the \dSST satisfy eq. (\ref{iias}) ($\tau$ is the proper time in both the $M^5$ and the \dSST). Taking the $\tau$ derivative, we get
 \beq X\^A\eta\_{AB}\dot X\^B=0.  \eeqno{masa}
So  $n\^A=X\^A/|X|=X\^A/\dsr$ -- the unit radial vector in the $M^5$ at $X\^A$ -- is perpendicular to all tangent vectors to the \dSST at $X\^A$; i.e., it is the local unit normal to the \dSST.
\par
Differentiate eq. (\ref{masa}) we get
\beq\dot X\^A\eta\_{AB}\dot X\^B+X\^A\eta\_{AB}\ddot X\^B=0.  \eeqno{masba}
Since $\dot X\^A\eta\_{AB}\dot X\^B=-1$  we have from eq. (\ref{masba})
\beq  n\^A\eta\_{AB}\ddot X\^B=1/\dsr=\alam. \eeqno{mufa}
$a\_5\^A=\ddot X\^A$ is the covariant acceleration on $\WW$ viewed as a worldline in the $M^5$.
We thus found that the component of $a\_5\^A$ in the radial direction is always the constant $\alam$ for any observer in the \dSST.
\par
$a\_5\^A$ is not the covariant acceleration on $\WW$ viewed as a worldline in the \dSST, which is
$ a^\m=D^2x^\m/D\tau^2=\ddot x^\m+\con{\m}{\n\a}\dot x^\n\dot x^\a$,
with $\con{\m}{\n\a}$ the connection in the \dSST. So, for example, a freely falling (geodesic) observer in the \dSST has $a^\m=0$, but $a\_5\^A=\alam n\^A$.
Writing $a^2=\gmn a^\m a^\n$, where $\gmn$ is the metric in the \dSST , it is straightforward to show that, in accordance with eq. (\ref{mufa}), everywhere on any world line in the \dSST
 \beq a\_5^2=a^2+\alam\^2. \eeqno{nuta}
\par
In summary, \dSST is a (pseudo) sphere of a fixed radius $\dsr$ where every worldline has an acceleration component $\alam=1/\dsr$ in the radial direction in the $M^5$ embedding ST, resonating with the discussion of the previous sections.
\par
This embedding and the accompanying facts are usually considered just a descriptive device. In the present context we should view this embedding as real with associated dynamics in the embedding ST which gives rise to the dynamics observed within the \dSST. For example, we would view the \dSST as subject to a radial potential field $\fm(\xi)$
where $\xi$ is the radius in $M^5$. The action of $\fm$ on the \dSST balances tension of the \dSST, part of which is the contribution of the possibly very large vacuum energy density. This balance occurs at some radius $\dsr$, thus making the effect of the large vacuum energy `disappear' and so eliminating the `old CC problem'. The value of $\dsr$ at which balance occurs is very large compared with the `natural' value for vacuum energy -- e.g. the Planck length -- and is observed as a small CC $\Lambda=3\dsr^{-2}$.
\par
The constant radial acceleration $\alam$ in $\E$ may seem as an artifact of no dynamical significance for matter in the \dSST. But in fact it can be said to have real observable effects within the \dSST via the Gibbons-Hawking effect and via its modification of the Unruh temperature of accelerated observers in the \dSST (see the discussion of these in Ref. \refcite{milgrom99} with references therein).
\par
Extending the ideas in previous sections to \dSST would require some serious generalizations: For example, \dSST is not compact and would require treating contributions from the boundaries (at $t=\pm\infty$). These boundaries are  Euclidean 3-spheres such as we have treated. Also, integrals over space which have appeared in our treatment will have to be generalized to ST integrals, introducing not only space nonlocality, but also time nonlocality, which, however, is not a stranger to MOND \cite{milgrom94a}.
\section{Discussion}
\label{discussion}
Even the simplistic model discussed above already offers important insights on the interpretation of various aspects of gravity in general and of MOND in particular. This includes hints on how to extend MOND.
For example, if the gravitational potential is but the distance from some center in $\E$ then it may give natural motivation for taking $\az$ to depend on $\f$, or even more general extensions. In fact, this approach shows us how $\az$ can altogether lose its role as a universal constant in the fully relativistic regime.
\par
Some of the important elements that went into the model are as follows: (a) Our brane Universe and matter on it,  couple to a potential $\fm(\xi)$ that acts in the embedding space, $\E$. In the example model I discussed, the force acts in the radial, $\vU$ direction (along the extra coordinate $\xi$) from some preferred center in $\E$. (b) The brane is also subject to a local moment that pulls its normal to local alignment with $\vU$. This is required in order to get MOND dynamics. (c) It is required that matter, while it can move with the brane in the $\vU$ direction,  has no inertia in this direction: matter inertia is not isotropic in $\E$ but is confined to the hyperplane perpendicular to $\vU$. Inertia of some degree of freedom is encapsuled in the free (kinetic) action; so the statement implies that only motion tangent to the local average brane enters the free action of matter. (d) The global brane configuration (e.g., a near sphere or pseudo-sphere of radius $\xi\approx\lz$) is determined by a balance between the brane tension and action of the external force on the brane. Local perturbations on the globally spherical brane are produced by a balance between the brane energy and the (radial) force on matter.
\par
What we can get out from such a picture -- worked out only in the `NR' limit corresponding to small local departures of the brane radius from the constant $\lz$ -- is as follows: (a) Nonrelativistic, gravitational dynamics involving masses that may reproduce MOND behaviors. (b) The MOND constant $\az$ is naturally introduced as the value of the $\fm$ force per unit mass at the brane radius: $\az\sim \fm'(\lz)$. (b) The gravitational potential $\f$ is proportional to the (small) local departure $\xi-\lz$ of the brane from a sphere. (c) The gravitational acceleration (gradient of the gravitational potential) measures the angle $\psi$ between the local normal to the brane and $\vU$. $|\gf|=\az$ corresponds to $\tan{\psi}\sim 1$. Thus the deep-MOND limit corresponds to very small departures of the brane normal from the radial, and the Newtonian regime corresponds to large departures. A brane Lagrangian that reproduces the exact MOND behavior that is required still has to be invoked, rather than derived from microscopics of the brane. (d) Balance of the brane, as in (d) above, implies that $\az\sim c^2/\lz$, where $c$ is the velocity of perturbations; thus suggesting the origin of the observed MOND-cosmology `coincidence' (\ref{coinc}).
(e) With this relation and the meaning of $c$, NR gravity $\f-\f\_{\infty}\ll c^2$ corresponds to small departures of the brane from sphericity, $|\xi-\lz|\ll\lz$. (The requirement from matter velocities $|\vv|\ll c$ also underlies the NR limit.) (f) The model points to a possible way out of the `old' cosmological-constant problem, in that even a very large, but position independent contribution to the brane tension -- such as a CC-like contribution -- is automatically balanced out by the $\fm'$ field by having the brane adjust its radius $\lz$. This occurs at a `large' radius -- large compared with the scales relevant to particle physics. The equilibrium radius is then perceived by us as a `small' cosmological constant $\dsr^{-2}\sim\lz^{-2}$. To actually base a solution on this argument it will be necessary to generalize our discussion of the brane as a space-like sphere to a brane as a space-time pseudosphere. Initial look at \dSST gives hope that this will be possible. Indeed, much remains to be elaborated on and understood perhaps from more basic principles. Almost every aspect of the model requires further elaboration.
But even as it is, it seems to me that we already get out of the model more than we have put in.

\end{document}